\documentstyle[12pt]{article}

\begin{document}

\centerline{\bf Nuclear Multifragmentation Critical Exponents}

\vspace*{1cm}

In a recent Letter \cite{EoSPRL} the EoS collaboration presented data of
fragmentation of 1 A~GeV gold nuclei incident on carbon.  By analyzing
moments of the fragment charge distribution, the authors claim
to determine the values of the critical exponents $\gamma$, $\beta$,
and $\tau$ for finite nuclei.  These data represent a crucial step forward in
our understanding of the physics of nuclear fragmentation.  However, as we
will show in the following, the analysis presented in \cite{EoSPRL} is not
sufficient to support the claim that the critical exponents for nuclear
fragmentation have been unambiguously determined.

The main difficulty  in observing critical behavior in
nuclear fragmentation is that nuclei
cannot be prepared and held near the temperature and density associated with
the critical point. Instead, complicated nuclear reactions must be used to
excite the nuclei, which may then expand to conditions sufficiently close to
the critical point.  The time spent at these conditions during the reaction
is an open question. However, even if
one assumes that conditions sufficiently close
to critical are explored in some of the observed
reactions, there remain at least two problems
with interpreting the subsequent experimental signals. One is to identify
which of the resulting particles have participated in the
equilibrated system near the critical conditions, and which have resulted from
the ``pre-equilibrium'' stage of the reaction. The other
problem is to measure the ``temperature'' of the decaying system. These
two problems are interrelated in the procedure used in Ref.~\cite{EoSPRL},
where the authors assume that
the observed multiplicity can be used as an indicator of temperature.
In this comment, we use the percolation model of nuclear
fragmentation \cite{Bau} to demonstrate the nature of these problems
in determining critical exponents.

In percolation models one uses a bond
breaking parameter $p$ with values between 0 and 1, including
$p_c$ , the ``critical value''.  In this model, near the critical point,
the charge of the largest
fragment is $Z_{\rm max} \propto (p-p_c)^\beta$, and the
second moment of the mass distribution is $M_2 \propto |p-p_c|^{-\gamma}$,
with $\beta=0.41$ and $\gamma=1.8$.

We follow the analysis of \cite{EoSPRL} and use $m_c$ = 26, in accordance with
the cut employed in \cite{EoSPRL}.  It is worth noting that we also find
roughly identical values of $\gamma$ and $\gamma'$ for the liquid and gas
branches, if we examine ln$M_2$ vs.\ ln$|m-m_c|$ and use $m_c$ = 26.
However, this
value of $m_c$ = 26 is lower than $m(p_c)$, and the numerically extracted
value of $\gamma$ = $\gamma'$ is approximately 1.2 to 1.4, significantly
lower than 1.8, the critical exponent for the percolation model.  These
two observations already indicate possible problems in trying to find the
critical exponents via the analysis employed in \cite{EoSPRL}.

One would only expect
$Z_{\rm max} \propto (m-m_c)^\beta$ and $M_2 \propto |m-m_c|^{-\gamma}$,
if multiplicity and $p$ are related in a strictly
linear way. Percolation calculations show, however, that on average
$m$ rises monotonically with $p$, but not linearly near the critical value.
Furthermore, for a given
value of $p$, there is always a distribution in the values of $m$.
Thus, any translation of an observable, $O$,
in terms of $p$ into terms of $m$ involves a non-trivial convolution,
$O(m) = \int dp\,m(p) \otimes O(p)$.

The results in Fig.~1 illustrate the difficulty in extracting the critical
exponent $\beta$.  Here we plot ln$Z_{\rm max}$ vs.\
ln$|m-m_c|$.  The open circles represent the result from
a percolation model with $Z$ = 79 charges.  $10^4$ events were generated
to provide statistics comparable to the experimental data.
The best fit to these points results in a slope `$\beta$' = 0.55.
For comparison,
the solid line has a slope of $\beta=0.41$, the nominal value for
percolation.  This demonstrates that the slope extracted from the present
plot is not the critical exponent $\beta$.

In addition to this problem,
one expects that a portion of the observed
multiplicity is comprised of pre-equilibrium particles, which are
not part of the equilibrated system.
In cascade simulations, the number of such charges varies
from 0 to possibly 15 for the system used in Ref.~\cite{EoSPRL}.
If it is not possible to experimentally (via momentum space
analysis, for example) remove these, then complicated modelling of the
pre-equilibrium reaction dynamics is needed.
To illustrate the influence of such pre-equilibrium contamination on
the extraction of meaningful values for the critical exponent $\beta$
we show the results of a calculation in which we assume that 10
pre-equilibrium particles exist along with an equilibrated system of 69
charges.  The remaining system of 69 charges is then
fragmented, again by using the percolation model.  Since the total
observed multiplicity
includes the pre-equilibrium particles, plotting ln$Z_{\rm max}$ vs.\
ln$|m-m_c|$ results in the crosses in Fig.~1.  For comparison, the dashed line
has a slope of `$\beta$' = 0.29.  (Coincidentally, this is the value
extracted from the experimental data.)

Similarly, when following the steps of analysis in \cite{EoSPRL} for
determination of $\gamma$, we find significant contamination of the `liquid'
branch due to pre-equilibrium emission.  In particular, we find that the
extremely large offset in ln$|M_2|$ between the liquid and gas branch (about
3.5) in Fig.~2 of \cite{EoSPRL} seems to be caused by this contamination,
in combination with the procedure of dropping the largest fragment on
the ``liquid'' side of the critical point and keeping it on the ``gas'' side.

Since the analysis of \cite{EoSPRL} does not work for a simple model
with known values of critical exponents, one should not expect it to yield
the correct critical exponents for the data, either.

In summary, while we applaud the effort and beautiful data of \cite{EoSPRL},
we do not agree with the conclusion that the critical indices of nuclear
fragmentation have been determined yet.

\vspace*{1cm}

\noindent Wolfgang Bauer$^{\dag}$ and William A. Friedman$^{\S}$

Institute for Nuclear Theory

University of Washington

Seattle, WA 98195, USA

\noindent
$^{\dag}$ Permanent address:

National Superconducting Cyclotron Laboratory and

Department of Physics and Astronomy, Michigan State University

East Lansing, MI 48824, USA

\noindent
$^{\S}$ Permanent address:

Department of Physics, University of Wisconsin,

Madison, WI 53706, USA

\vspace*{0.4 cm}

\clearpage

\vspace*{10cm}

\includegraphics{Fig1.ps}

\vfill
\noindent {\bf Fig.\ 1}.  Percolation model simulation of the fragmentation
of gold.  Circles:  $Z=79$ system fragmenting; crosses: $Z=69$ system
fragmenting plus 10 pre-equilibrium protons.  For comparison,
the solid line has a
slope of $\beta$ = 0.41, and the dashed line `$\beta$' = 0.29.

\end{document}